\documentclass[aps,prb, amsmath,amssymb,reprint,graphicx,showpacs,superscriptaddress]{revtex4-1}

\usepackage{graphicx}
\usepackage{dcolumn}
\usepackage{bm}

\begin{document}


\title{Spin-dependent current through a quantum dot from spin-polarized non-equilibrium quantum Hall edge channels }


\author{H. Kiyama}
\email{kiyama@sanken.osaka-u.ac.jp}
\affiliation{The Institute of Scientific and Industrial Research, Osaka University, 8-1, Mihogaoka, Ibaraki-shi, Osaka 567-0047, Japan}
\author{T. Nakajima}
\affiliation{Center for Emergent Matter Science, RIKEN, 2-1 Hirosawa, Wako-shi, Saitama 351-0198, Japan}
\author{S. Teraoka}
\affiliation{Department of Applied Physics, The University of Tokyo, 7-3-1 Hongo, Bunkyo-ku 113-8656, Japan}
\author{A. Oiwa}
\affiliation{The Institute of Scientific and Industrial Research, Osaka University, 8-1, Mihogaoka, Ibaraki-shi, Osaka 567-0047, Japan}
\affiliation{Department of Applied Physics, The University of Tokyo, 7-3-1 Hongo, Bunkyo-ku 113-8656, Japan}
\author{S. Tarucha}
\affiliation{Center for Emergent Matter Science, RIKEN, 2-1 Hirosawa, Wako-shi, Saitama 351-0198, Japan}
\affiliation{Department of Applied Physics, The University of Tokyo, 7-3-1 Hongo, Bunkyo-ku 113-8656, Japan}

\date{\today}

\begin{abstract}
We report selective injection of both spin-up and spin-down single electrons into a quantum dot (QD) from spin-polarized non-equilibrium quantum Hall edge channels (ECs) generated by selective transmission of spin-resolved ECs using a surface gate placed at a distance from the QD.  We change the spin polarization of non-equilibrium ECs by changing the bias voltages applied to different source Ohmic contacts. The efficiency of spin-up electron injection reaches 0.5, which is approximately 0.2 higher than that induced by spin-dependent tunnel coupling between QD and ECs. On the other hand, the efficiency of spin-down electron injection reaches 0.4. In addition, we rectify the underestimation of the efficiency of spin filtering for equilibrium ECs by numerically subtracting the contribution of the excited states in the QD. The obtained spin-filtering efficiency is higher than that evaluated from the raw experimental data and increases with magnetic field as expected with the increase in the spatial separation between ECs.
\end{abstract}


\maketitle 
\section{Introduction}
Quantum dots (QDs) formed in a two-dimensional electron gas (2DEG) are increasingly gaining attention for their applications in spintronics and spin-based quantum information processing. In particular, preparation and detection of electron spin states in QDs are key ingredients of the recent progress in coherent spin manipulation\cite{Petta,Nowack,Michel,Medford} and device scalability.\cite{takakura} In an important technique, spin filtering has been reported for a QD contacted by spatially spin-resolved quantum Hall edge channels (ECs).\cite{CiorgaPRB,Rogge1,Otsuka,Kiyama} The ECs usually have stronger tunnel coupling to the QD for lower energy spin (spin-up) electrons than for higher energy spin (spin-down) electrons. This spin-dependent tunnel coupling enables spin injection such that spin-up electrons predominantly tunnel into the QD rather than spin-down electrons from equilibrium ECs (EC spin filtering). However, injection of spin-down electrons is not feasible by this type of spin filtering. Injection of both spin-up and spin-down electrons has been realized in self-assembled QDs \cite{Hamaya1,Hamaya2} and carbon nanotube QDs \cite{Sahoo,Jensen} with ferromagnetic contacts. In these devices, the spin orientation to be injected is switched only by reversing the magnetization of the ferromagnetic contacts, which is achieved by controlling the external magnetic field; electrical switching of the spin orientation has not yet been reported. 

The efficiency of the EC spin filtering, which corresponds to the spin polarization of the current through the QD, observed till date is subject to certain restrictions. In principle, the EC spin-filtering efficiency should increase with the magnetic field because spin-up and spin-down ECs are more spatially separated from each other. However, it has been reported in an experiment using a spin-polarized quantum wire instead of ECs that the spin-filtering efficiency increases with an increasing magnetic field in the low field range and decreases in the higher field range.\cite{Hitachi} The origin of the decreased efficiency is supposedly the result of electron conduction through the excited states of the QD, but this has not been investigated experimentally yet.

In this paper, we demonstrate selective injection of both spin-up and spin-down single electrons from spin-polarized non-equilibrium ECs into a QD formed in a 2DEG. The spin-polarized non-equilibrium ECs are generated using a surface gate placed a few $\mu$m away from the QD. The spin polarization of non-equilibrium ECs and consequently that of the QD current are electrically switched between spin-up and spin-down by changing the bias voltages applied to source Ohmic contacts without reversing the external magnetic field. In addition, we observe spin-independent components in the QD current, which suggest an excess current passing through the excited states of the QD. By numerically eliminating this contribution of the excited states, we rectify the underestimated efficiency of EC spin-filtering.

The remainder of this paper is organized as follows. In Sec. II, we describe our device and the results of QD current measurements with both spin-up and spin-down polarized non-equilibrium ECs, demonstrating the selective injections of spin-up and spin-down electrons. In Sec. III, we evaluate the current through the excited states of the QD and subsequently the EC spin-filtering efficiency by eliminating the excited state contributions. Finally, we provide our conclusion in Sec. IV.

\section{Experiment with spin-polarized non-equilibrium edge channels}

Our device is fabricated from a GaAs/AlGaAs heterostructure with a 2DEG located 100 nm below the surface. The 2DEG has an electron density of $3\times10^{11} \mathrm{cm}^{-2}$ and a mobility of $1\times10^{6} \mathrm{cm^{2}/Vs}$. Figure 1(a) shows a scanning electron micrograph of the device. A single QD is formed by negatively biasing the Ti/Au Schottky gates L, P, R, and T. Gate LC is used to reduce the number of ECs beneath and to generate non-equilibrium current through the ECs. Gates colored in green completely deplete the 2DEG beneath. The distance between gate LC and the QD is approximately 2 $\mu$m. The size of gate LC is approximately 0.5 $\mu$m $\times$ 100 $\mu$m.

The measurements are performed with the device placed in a dilution refrigerator with a base temperature of 130 mK and an electron temperature of 350 mK. A magnetic field $B$ is applied with a tilt of $60^\circ$ from the normal to the 2DEG plane for enhancing Zeeman splitting at appropriate filling factors. The current through the QD is measured at the drain Ohmic contact [denoted as D in Fig. 1(a)] with bias voltages $V_{\mathrm{S1}}$ and $V_{\mathrm{S2}}$ applied to the Ohmic contacts S1 and S2, respectively. In QD current measurements, the perpendicular component $B_{\perp}$ of $B$ is changed between 1 and 2 T, corresponding to filling factors of 12 and 6, respectively. 

\begin{figure}
\includegraphics{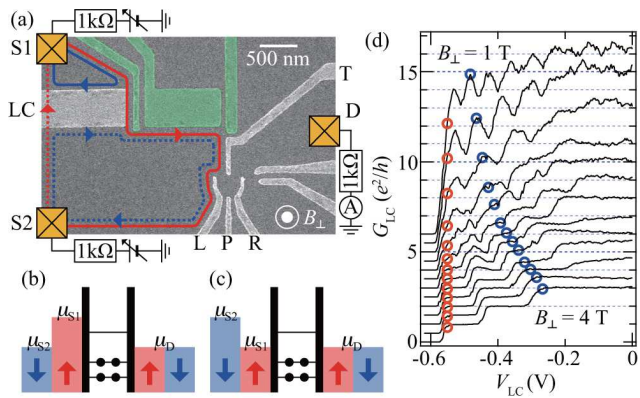}
\caption{(a) Scanning electron micrograph of the device with a schematic of the experimental
setup. Spin-up and spin-down edge channels in the lowest Landau level are drawn as red and blue lines, respectively. (b), (c) Energy diagrams with spin-up polarized (b) and spin-down polarized (c) non-equilibrium edge channels. (d) Conductance beneath gate LC as a function of gate voltage $V_{\mathrm{LC}}$ at different magnetic fields $B_{\perp}$.  $B_{\perp}$ is varied from 1 to 4 T with a 0.25 T step. Each curve is offset by 0.5$e^{2}/h$ for clarity. Red and blue circles show the left edges of the $e^{2}/h$ and $3e^{2}/h$ plateaus, respectively.
}%
\end{figure}

The red and blue lines in Fig. 1(a) show a schematic of the outermost spin-up and spin-down ECs, respectively, in the case in which only a spin-up EC is transmitted through gate LC.  The inner ECs in higher Landau levels are not shown since they can be ignored because of negligibly small tunnel coupling with the QD. The chemical potential of the outermost spin-up EC near the QD is defined by $V_{\mathrm{S1}}$, while that of the spin-down EC is defined by $V_{\mathrm{S2}}$. In the experiment with spin-up polarized non-equilibrium ECs, chemical potentials $\mu_{\mathrm{S1}}$, $\mu_{\mathrm{S2}}$, and $\mu_{\mathrm{D}}$ at Ohmic contacts S1, S2, and D, respectively, are set to satisfy the condition that $\mu_{\mathrm{S1}} > \mu_{\mathrm{S2}} = \mu_{\mathrm{D}}$ [Fig. 1(b)]. In the case of spin-down polarized non-equilibrium ECs, they are changed to satisfy the condition that $\mu_{\mathrm{S2}} > \mu_{\mathrm{S1}} = \mu_{\mathrm{D}}$ [Fig. 1(c)]. 

Figure 1(d) shows the two-terminal conductance through gate LC, $G_{\mathrm{LC}}$, as a function of the gate voltage $V_{\mathrm{LC}}$ on gate LC at different perpendicular magnetic fields $B_{\perp}$. $G_{\mathrm{LC}}$ oscillates with $V_{\mathrm{LC}}$ for $B_{\perp} <$ 3 T, and shows quantized plateaus for higher $B_{\perp}$. These oscillations are attributed to the scattering between the counter-propagating ECs at opposite side of gate LC via localized states under gate LC, which emerges because of potential fluctuations when each Landau level is energetically close to the Fermi level.\cite{Haug2} Red and blue circles show the left edges of the $e^{2}/h$ and $3e^{2}/h$ plateaus, respectively. Though the plateaus are not clearly observed in the low $B_{\perp}$ range, we mark the blue circles by linear extrapolation from the data in the higher $B_{\perp}$ range according to the linear $B_{\perp}$ dependence of the state degeneracy for each Landau level, while we assume a constant gate voltage for the red circles. 

	We first use an EC spin-filtering technique to assign the ground spin states in the QD. Figure 2 shows the intensity plot of the conductance through the QD, $G_{\mathrm{QD}}$, measured as a function of the gate voltage  $V_{\mathrm{P}}$ and  $B_{\perp}$ with a small bias voltage of 30 $\mu$V applied across the QD. The ECs are set in equilibrium by completely depleting the 2DEG beneath gate LC. The Coulomb peaks show a zigzag structure, indicating crossings of the lowest Landau level (LL0) and the second lowest Landau level (LL1).\cite{McEuen} The LL0 ridges clearly show the alternating conductance intensity due to alternating spin state in the QD and the effect of EC spin filtering. At the high conductance LL0 ridges, electrons tunneling through the QD are assigned to be spin-up, while they are spin-down at the LL0 ridges with low conductance. These features are shown schematically in the bottom of Fig. 2. The electron number in the QD is roughly estimated to be between 10 and 20 from the charging energy of approximately 1.3 meV and the magnetic field value for the QD filling factor $\nu_{\mathrm{QD}}$ = 2 shown by a white dashed line in Fig. 2. 

\begin{figure}
\includegraphics{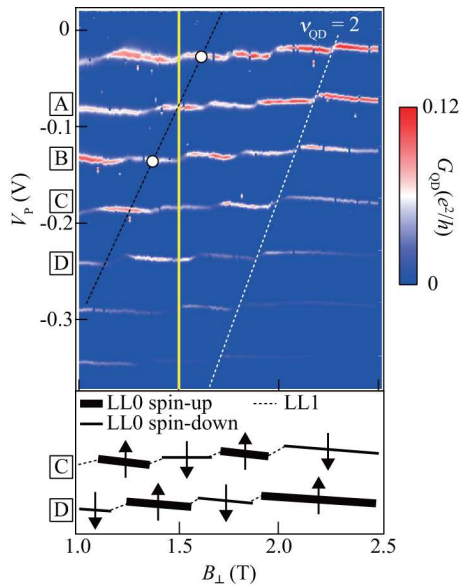}
\caption{(Color online)(Top) Intensity plot of the QD conductance $G_{\mathrm{QD}}$ with the ECs in equilibrium as a function of the perpendicular magnetic field $B_{\perp}$ and the gate voltage $V_{\mathrm{P}}$. The QD filling factor $\nu_{\mathrm{QD}}$ = 2 is shown as a dashed line. $G_{\mathrm{QD}}$ shown as open circles is used to estimate spin-down conductance in the electrostatic condition at peak A in $B_{\perp}$ = 1.5 T. (Bottom) Schematic for assigning spins tunneling through the QD. The thick solid, thin solid, and dashed lines show LL0 ridges with high conductance, LL0 ridges with low conductance, and LL1 ridges, respectively. The up and down arrows indicate spin orientation.}%
\end{figure}

Next, we describe the QD current measurement using non-equilibrium ECs. Figure 3(a) shows the QD current at Coulomb peaks labeled A and C in Fig. 2 as a function of $V_{\mathrm{LC}}$ at $B_{\perp}$ = 1.5 T. $V_{\mathrm{S1}}$ and $V_{\mathrm{S2}}$ are set to --60 $\mu$V and 0 $\mu$V, respectively, for spin-up polarized non-equilibrium ECs. Both peaks almost reach zero around $V_{\mathrm{LC}} = -0.59$ V, reflecting the complete depletion of the 2DEG beneath gate LC, though a small current remains probably because of a source bias voltage fluctuation. At $B_{\perp} = 1.5$ T (shown by the yellow line in Fig. 2), electrons conducting through the QD are assigned to be spin-up for peak A, and spin-down for peak C. 

In Fig. 3(a), for peak A, the QD current starts to decrease as $V_{\mathrm{LC}}$ decreases downward from $V_{\mathrm{LC}} = -0.55$ V ($\equiv V_{\alpha}$) and from $-0.44$ V ($\equiv V_{\delta}$). These two gate voltages are matched to the left edges of the $e^{2}/h$ and $3e^{2}/h$ plateaus in  $G_{\mathrm{LC}}$ as shown by the red and blue circles, respectively, in Fig. 1(d). In contrast, though the QD current for peak C also shows a similar $V_{\mathrm{LC}}$ dependence, this peak shows decreases at $V_{\mathrm{LC}} = -0.51$ V ($\equiv V_{\beta}$) and at $-0.40$ V ($\equiv V_{\varepsilon}$). These two gate voltages are slightly larger than those for peak A, and are presumably assigned to the left edges of the $2e^{2}/h$ and $4e^{2}/h$ plateaus in $G_{\mathrm{LC}}$, respectively. These results imply that the spin-resolved local filling factor beneath gate LC can be detected using a distant QD. However, the changes in the QD current at $V_{\delta}$ and $V_{\varepsilon}$ cannot be attributed to the change in the number of electrons tunneling directly from the inner ECs into the QD because the QD dominantly couples to the outermost spin-up and spin-down ECs and the couplings to the inner ECs are negligibly small. 

\begin{figure}
\includegraphics{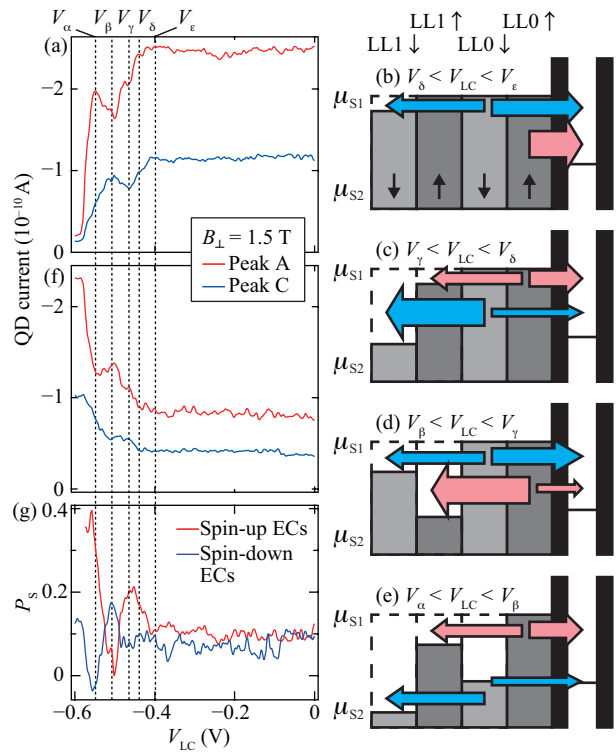}
\caption{(Color online)(a)$V_{\mathrm{LC}}$ dependence of the QD current at Coulomb peaks A and C in $B_{\perp}$ = 1.5 T with $V_{\mathrm{S1}} = -60$ $\mu$V and $V_{\mathrm{S2}} = 0$ $\mu$V for spin-up non-equilibrium ECs. (b)-(e) Schematics to explain the chemical potentials of ECs at different $V_{\mathrm{LC}}$. Left and right arrows represent the electron transfer between ECs without spin-flip and the electron conduction through the QD, respectively, with a red (blue) color for spin-up (spin-down). Their thickness denotes the number of electrons flowing to the QD or the inner ECs. (f) Same as (a) but with $V_{\mathrm{S1}} = 0$ $\mu$V and $V_{\mathrm{S2}} = -60$ $\mu$V for spin-down non-equilibrium ECs. (g) Spin polarization $P_{\mathrm{S}}$ for peak A in $B_{\perp}$ = 1.5 T for spin-up (red) and spin-down (blue) non-equilibrium ECs. Vertical dashed lines $V_{\alpha}$,  $V_{\beta}$, $V_{\gamma}$, $V_{\delta}$ and $V_{\varepsilon}$ are guide lines for $V_{\mathrm{LC}}$ = --0.55 V, --0.51 V, --0.46 V, --0.44 V, and --0.4 V, respectively.}%
\end{figure}

We now describe a possible origin for the change in the QD current at $V_{\delta}$ and $V_{\varepsilon}$ that is schematically explained in Fig. 3(b) to 3(e), together with the reason for the slight increase in the QD current with decreasing $V_{\mathrm{LC}}$ from $V_{\mathrm{LC}} = V_{\beta}$ ($V_{\mathrm{LC}} =-0.46$ V $\equiv V_{\gamma}$) for peak A (peak C). In the region $V_{\delta} < V_{\mathrm{LC}} <V_{\varepsilon}$, the second outermost spin-down EC (EC$_{\mathrm{LL1}\downarrow}$) is partially transmitted through gate LC. The energy distribution of the transmitted EC is given as  $f(E) = Tf_{\mathrm{S1}}(E) + (1-T) f_{\mathrm{S2}}(E)$,\cite{Altimiras} where $f_{\mathrm{S1}}$ and $f_{\mathrm{S2}}$ are the equilibrium Fermi distribution functions in the EC$_{\mathrm{LL1}\downarrow}$ for the chemical potentials $\mu_{\mathrm{S1}}$ and $\mu_{\mathrm{S2}}$, respectively, and $T$ is the transmission probability for EC$_{\mathrm{LL1}\downarrow}$ through gate LC. Though the non-equilibrium energy distribution in a single EC, $f(E)$, may be unchanged over a few $\mu$m,\cite{Altimiras,leSueur} there are no signatures of deviation from the equilibrium distribution in our experiments such as the broadening of Coulomb peaks, possibly because of the high electron temperature of 350 mK. Therefore, for simplicity, we assume an equilibrium distribution in EC$_{\mathrm{LL1}\downarrow}$ after the transmission through gate LC with a chemical potential between $\mu_{\mathrm{S1}}$ and $\mu_{\mathrm{S2}}$ as illustrated in Fig. 3(b), while there remains a non-equilibrium distribution between different ECs. 

During the propagation of 2 $\mu$m from gate LC to the distant QD, electrons in the outermost spin-down EC (EC$_{\mathrm{LL0}\downarrow}$) are transferred to EC$_{\mathrm{LL1}\downarrow}$ by impurity and phonon scatterings without spin-flip as illustrated in Fig. 3(b). Thus, in the region $V_{\delta} < V_{\mathrm{LC}} <V_{\varepsilon}$, EC$_{\mathrm{LL0}\downarrow}$ is depopulated, and consequently the QD current decreases for peak C, which is a spin-down LL0 ridge at $B_{\perp}$ = 1.5 T. Notably, the equilibration length for these ECs is approximately a few $\mu$m in low magnetic fields.\cite{Hirai} We expect that the electron transfer from the spin-up ECs to EC$_{\mathrm{LL1}\downarrow}$ is less efficient because of the need for a spin-flip process.\cite{Muller} Also, we neglect the electron transfer from EC$_{\mathrm{LL1}\downarrow}$ to the inner spin-down ECs for simplicity. 

In the region $V_{\gamma} < V_{\mathrm{LC}} < V_{\delta}$, the QD current similarly decreases for peak A, which is a spin-up LL0 ridge at $B_{\perp}$ = 1.5 T, because of the electron transfer from the outermost spin-up EC (EC$_{\mathrm{LL0}\uparrow}$) to the second outermost spin-up EC (EC$_{\mathrm{LL1}\uparrow}$), as shown in Fig. 3(c). The QD current for peak C further decreases from that in the region $V_{\delta} < V_{\mathrm{LC}} < V_{\varepsilon}$ because EC$_{\mathrm{LL0}\downarrow}$ becomes depopulated more efficiently as the chemical potential of EC$_{\mathrm{LL1}\downarrow}$ is lowered.

In the region $V_{\beta} < V_{\mathrm{LC}} < V_{\gamma}$, the QD current for peak C increases with decreasing $V_{\mathrm{LC}}$, while that for peak A keeps decreasing in the same way as in the region $V_{\gamma} < V_{\mathrm{LC}} < V_{\delta}$. The former result suggests that the inter-EC electron transfer described above is suppressed in this $V_{\mathrm{LC}}$ range. We believe that this suppression is caused by the increase in the chemical potential of EC$_{\mathrm{LL1}\downarrow}$. When the spin-down LL0 under gate LC is close to the Fermi energy, the electrons in EC$_{\mathrm{LL1}\downarrow}$ can be transmitted through gate LC via localized states,\cite{Haug2} as implied in Fig. 1(d). Therefore, the chemical potential of EC$_{\mathrm{LL1}\downarrow}$ becomes higher toward $\mu_{\mathrm{S1}}$, as the spin-down LL0 approaches the Fermi energy with decreasing $V_{\mathrm{LC}}$ as illustrated in Fig. 3(d). Consequently, the electron transfer from EC$_{\mathrm{LL0}\downarrow}$ to EC$_{\mathrm{LL1}\downarrow}$ is suppressed because the difference in the chemical potentials between these ECs decreases.

Finally, in the region $V_{\alpha} < V_{\mathrm{LC}} < V_{\beta}$, the QD current for peak A increases as the transmission of EC$_{\mathrm{LL1}\uparrow}$ through gate LC increases with decreasing $V_{\mathrm{LC}}$ in the same way as for peak C in the region $V_{\gamma} < V_{\mathrm{LC}} < V_{\delta}$ but for spin-up. For peak C, the QD current decreases with decreasing $V_{\mathrm{LC}}$ because EC$_{\mathrm{LL0}\downarrow}$ is partially transmitted under gate LC and then electrons in EC$_{\mathrm{LL0}\downarrow}$ are scattered to EC$_{\mathrm{LL1}\downarrow}$ during the propagation between gate LC and the QD. 

The electron transfer between ECs with the same spin orientation described in Fig. 3(b) to 3(e) is expected to be suppressed in high magnetic fields since the overlap of wavefunctions between the different ECs is reduced. In our experiments, the magnitude of the change in the QD current such as that in $V_{\alpha}  < V_{\mathrm{LC}} < V_{\varepsilon}$ in Fig. 3(a) becomes small with increasing $B_{\perp}$ in the field range higher than 1.5 T, which agrees with the above expectation. However, the electron transfer seems to be suppressed with decreasing $B_{\perp}$ in the range lower than 1.5 T. This might be due to the small Landau level splitting at low $B_{\perp}$. In this case, the scattering between the counter-propagating ECs across gate LC occurs efficiently, as indicated by higher $G_{\mathrm{LC}}$ at lower $B_{\perp}$ for $V_{\mathrm{LC}}$ giving same filling factor under gate LC in Fig. 1(d). Therefore, EC$_{\mathrm{LL1}\uparrow (\downarrow)}$ and EC$_{\mathrm{LL0}\uparrow (\downarrow)}$ are kept close to equilibrium over $V_{\alpha}  < V_{\mathrm{LC}} < V_{\varepsilon}$, and the effect of the electron transfer may not appear to be significant at low $B_{\perp}$.

Next, we prepare the spin-down non-equilibrium ECs near the QD by setting the condition of the bias voltage applied to the Ohmic contacts such that $V_{\mathrm{S1}} = 0$ $\mu$V and $V_{\mathrm{S2}} = -60$ $\mu$V. Figure 3(f) shows the QD current for Coulomb peaks A and C as a function of $V_{\mathrm{LC}}$. For $V_{\mathrm{LC}} > -0.4$ V, there is a finite current due to the residual negative bias voltage at the Ohmic contact S1 because $V_{\mathrm{S1}}$ and $V_{\mathrm{S2}}$ are applied through the external circuit with 1 k$\Omega$ resistance, although the QD current should ideally be zero when all of the ECs are grounded by S1. For $V_{\mathrm{LC}}<$ --0.4 V, the features of the change in the QD current are very similar to those observed in Fig. 3(a), though the increases and decreases are reversed. For peak A, the QD current increases with decreasing $V_{\mathrm{LC}}$ in the regions $V_{\mathrm{LC}} < V_{\alpha}$ and $V_{\beta} < V_{\mathrm{LC}} < V_{\delta}$. This indicates that EC$_{\mathrm{LL0}\uparrow}$ and EC$_{\mathrm{LL1}\uparrow}$ biased at the Ohmic contact S2 are reflected at gate LC for $V_{\mathrm{LC}} < V_{\alpha}$ and for $V_{\beta} < V_{\mathrm{LC}} < V_{\delta}$, respectively. Electrons in EC$_{\mathrm{LL1}\uparrow}$ with an energy higher than the chemical potential in EC$_{\mathrm{LL0}\uparrow}$ can be transferred to EC$_{\mathrm{LL0}\uparrow}$ by inelastic scattering. Therefore, the reflection of EC$_{\mathrm{LL1}\uparrow}$ is manifested as an increase in the QD current. On the other hand, for peak C, the QD current increases with decreasing $V_{\mathrm{LC}}$ in the region $V_{\mathrm{LC}} < V_{\beta}$. This means that EC$_{\mathrm{LL0}\downarrow}$ biased at the Ohmic contact S2 is reflected at gate LC. A similar increase is expected when EC$_{\mathrm{LL1}\downarrow}$ is reflected at gate LC  with decreasing $V_{\mathrm{LC}}$ for $V_{\gamma} < V_{\mathrm{LC}} < V_{\varepsilon}$. However, no distinct increase is observed for peak C in this $V_{\mathrm{LC}}$ range. The reason for the absence of this increase is not yet clear.

We evaluate the spin polarization $P_{\mathrm{S}}$ of electrons conducting through a spin-degenerate QD, which is defined as 
\begin{equation}
P_{\mathrm{S}}=\frac{I_{\uparrow}-I_{\downarrow}}{I_{\uparrow}+I_{\downarrow}}
\end{equation}
where $I_{\uparrow}$ and $I_{\downarrow}$ are the spin-up and spin-down components of the QD current, respectively. Positive $P_{\mathrm{S}}$ (negative $P_{\mathrm{S}}$) means the injection of spin-up (spin-down) electrons into the QD. The spin-up component $I_{\uparrow}$ at the gating condition of peak A is given by the QD current at the LL0 spin-up ridge, $I_{\uparrow}^{\mathrm{A}}$, which is directly measured for peak A. On the other hand, the spin-down component $I_{\downarrow}$ at the same gating condition should be given by the QD current at the LL0 spin-down ridge. For this purpose, we use the QD current measured for peak C, $I_{\downarrow}^{\mathrm{C}}$, and compensate the difference in the tunnel barrier between the two gating conditions for peaks A and C. The value of $I_{\downarrow}$ at the gating condition of peak A is derived using $I_{\downarrow}^{\mathrm{C}}$ as $I_{\downarrow} = (I_{0\downarrow}^{\mathrm{A}}/I_{0\downarrow}^{\mathrm{C}} ) I_{\downarrow}^{\mathrm{C}}$, where $I_{0\downarrow}^{\mathrm{X}}$ is the spin-down QD current with equilibrium ECs at peak X (X = A, C). Because peak A is on the LL0 spin-up ridge, $I_{0\downarrow}^{\mathrm{A}}$ is approximately estimated by the average of the QD currents at two adjacent LL0 spin-down ridges shown by open circles in Fig. 2, while $I_{0\downarrow}^{\mathrm{C}}$ is the raw QD current for peak C. 

A red curve in Fig. 3(g) shows $P_{\mathrm{S}}$ with spin-up non-equilibrium ECs as a function of $V_{\mathrm{LC}}$. For $V_{\mathrm{LC}} > -0.4$ V, the non-equilibrium ECs near the QD are not spin-polarized; therefore, $P_{\mathrm{S}}$ is approximately 0.1, which corresponds to the efficiency of EC spin filtering. $P_{\mathrm{S}}$ changes with $V_{\mathrm{LC}}$ for $V_{\mathrm{LC}} < -0.4$ V, reflecting the changes in $I_{\uparrow}$ and $I_{\downarrow}$, and reaches 0.4 at $V_{\mathrm{LC}} \approx V_{\alpha}$. The tunnel barrier between the QD and ECs is independent of $V_{\mathrm{LC}}$ because gate LC is located far enough from the QD. Therefore, the increase in $P_{\mathrm{S}}$ due to the change in $V_{\mathrm{LC}}$ implies that the non-equilibrium ECs near the QD are spin-up polarized and injected into the QD. On the other hand, $P_{\mathrm{S}}$ measured for the spin-down non-equilibrium ECs is shown as a blue curve in Fig. 3(f). For $V_{\mathrm{LC}} > -0.4$ V, $P_{\mathrm{S}}$ is constant at 0.1, reflecting the EC spin-filtering efficiency. For $V_{\mathrm{LC}} < -0.4$ V, it varies in an opposite manner to $P_{\mathrm{S}}$ for the spin-up non-equilibrium ECs and decreases to --0.04 at $V_{\mathrm{LC}} \approx V_{\alpha}$. This implies that the spin-down QD current becomes large compared with the spin-up QD current.

\begin{figure}
\includegraphics{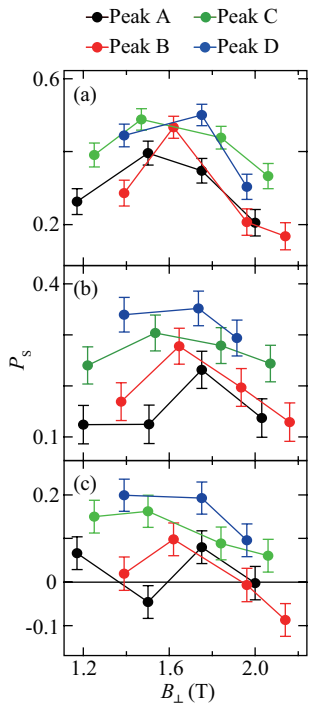}
\caption{(Color online)$B_{\perp}$ dependence of maximum $P_{\mathrm{S}}$ for spin-up non-equilibrium ECs (a), $P_{\mathrm{S}}$ for equilibrium ECs (b), and minimum $P_{\mathrm{S}}$ for spin-down non-equilibrium ECs (c) for Coulomb peaks A, B, C, and D. }%
\end{figure}

Figures 4(a), (b), and (c) show the maximum $P_{\mathrm{S}}$ measured for the spin-up polarized non-equilibrium ECs, $P_{\mathrm{S}}$ measured for the equilibrium ECs (EC spin-filtering efficiency), and the minimum $P_{\mathrm{S}}$ measured for the spin-down polarized non-equilibrium ECs, respectively. The $P_{\mathrm{S}}$ values are evaluated for different Coulomb peaks as a function of $B_{\perp}$. For all Coulomb peaks and across the whole range of $B_{\perp}$, $P_{\mathrm{S}}$ is in general larger for the spin-up non-equilibrium ECs and smaller for the spin-down non-equilibrium ECs by approximately 0.2 compared to the EC spin-filtering efficiency.

As an overall trend, $P_{\mathrm{S}}$ shown in Fig. 4(a) to 4(c) increases with $B_{\perp}$ up to 1.6 T and decreases at higher $B_{\perp}$. We believe that this decrease is due to the increase in electron conduction through excited states in the QD with increasing $B_{\perp}$.\cite{Hitachi} The energy splitting between the ground and excited states generally decreases with $B_{\perp}$. When it becomes comparable to the electron temperature, there will be a considerable amount of spin-up current through the excited states at LL0 spin-down ridges. This overestimate of $I_{\downarrow}$ in Eq. (1) results in an underestimate of $P_{\mathrm{S}}$, even when the ECs are fully spin-up polarized. The reason for the limited $P_{\mathrm{S}}$ for the spin-up polarized non-equilibrium ECs is also expected to be the contribution of the excited state conduction. We estimate this contribution in Sec. III.

The negative values of $P_{\mathrm{S}}$ obtained for the spin-down polarized non-equilibrium ECs imply that $I_{\downarrow}$ becomes large compared with $I_{\uparrow}$ in Eq. (1). However, although this result indicates that more spin-down electrons tunnel into the QD at $V_{\mathrm{LC}} \approx V_{\alpha}$ than for equilibrium ECs, the small amplitude $|P_{\mathrm{S}}|$ implies that the QD current may not be significantly spin-down polarized. As we discuss later, the conduction of spin-up electrons through the excited states supposedly occurs at the spin-down LL0 ridges when EC$_{\mathrm{LL0}\uparrow}$ is negatively biased. Because of the residual negative bias voltage at the Ohmic contact S1 and the electron transfer from EC$_{\mathrm{LL1}\uparrow}$ to EC$_{\mathrm{LL0}\uparrow}$, there will be such a spin-up excess current in $I_{\downarrow}$ for peak C at $V_{\mathrm{LC}} \approx V_{\alpha}$ in Fig. 3(f). Therefore, the pure spin-down QD current may be smaller than the observed $I_{\downarrow}$, and consequently, the actual $P_{\mathrm{S}}$ may be higher than the experimentally obtained values in Fig. 4(c).

\begin{figure}
\includegraphics{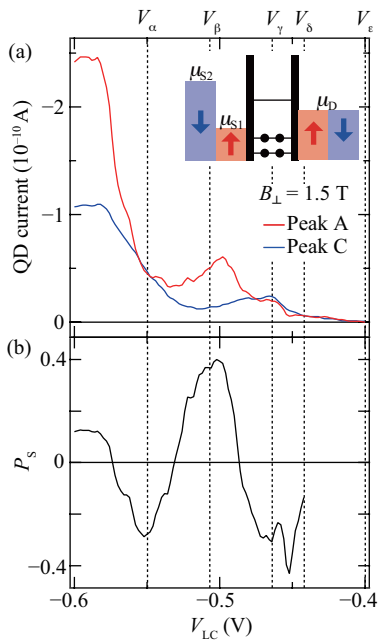}
\caption{(Color online)(a)$V_{\mathrm{LC}}$ dependence of the negative component of the QD current for Coulomb peaks A and C at $B_{\perp}$ = 1.5 T with  $V_{\mathrm{S1}} = +60$ $\mu$V and $V_{\mathrm{S2}} = -60$ $\mu$V for spin-down non-equilibrium ECs. (b) Spin polarization $P_{\mathrm{S}}$ for peak A calculated using the data shown in (a). (Inset) Energy diagrams with a positive (negative) bias voltage to the outermost spin-up (spin-down) EC for highly efficient injection of spin-down electrons into the QD.}%
\end{figure}

To suppress the spin-up excess current due to the residual negative bias voltage at the Ohmic contact S1, we  tune the chemical potential of EC$_{\mathrm{LL0}\uparrow}$ near the QD to be lower than $\mu_{\mathrm{D}}$ as illustrated in the inset of Fig. 5(a), by setting the bias voltages applied to the Ohmic contacts such that $V_{\mathrm{S1}} = +60$ $\mu$V and $V_{\mathrm{S2}} = -60$ $\mu$V. Under these conditions, the QD current is  negative (positive) when a QD level is located in between $\mu_{\mathrm{S2}}$ and $\mu_{\mathrm{D}}$ ($\mu_{\mathrm{S1}}$ and $\mu_{\mathrm{D}}$). In the negative QD current, there is no spin-up excess component from EC$_{\mathrm{LL0}\uparrow}$ as long as EC$_{\mathrm{LL0}\uparrow}$ is perfectly transmitted through gate LC. Figure 5(a) shows the negative QD current for this bias condition for Coulomb peaks A and C at $B_{\perp}$ = 1.5 T as a function of $V_{\mathrm{LC}}$. The QD current is almost zero at $V_{\mathrm{LC}} = V_{\varepsilon}$, indicating elimination of the residual negative bias voltage at the Ohmic contact S1. Besides a change in the QD current similar to that observed in Fig. 3(f) at $V_{\mathrm{LC}} = V_{\alpha}$, $V_{\beta}$, and $V_{\gamma}$, the increase in the QD current is observed with decreasing $V_{\mathrm{LC}}$ from $V_{\mathrm{LC}} = V_{\delta}$ for peak A and from $V_{\mathrm{LC}} = V_{\varepsilon}$ for peak C. Peak A also shows a slight increase in the QD current with decreasing $V_{\mathrm{LC}}$ from $V_{\mathrm{LC}} = V_{\varepsilon}$. We consider that this increase is the result of the conduction of the spin-down electrons through the excited states, which we neglect in the spin-up injection because of the small tunnel coupling for spin-down ECs. These features are consistent with the electron transfer from the inner ECs to the outermost EC.

We evaluate $P_{\mathrm{S}}$ for spin-down non-equilibrium ECs using the data shown in Fig. 5(a) as a function of $V_{\mathrm{LC}}$ in Fig. 5(b). Though $P_{\mathrm{S}}$ changes with $V_{\mathrm{LC}}$ in a similar way to the case in which $V_{\mathrm{S1}} = 0$ $\mu$V and $V_{\mathrm{S2}} = -60$ $\mu$V in Fig. 3(f), it reaches as low as $P_{\mathrm{S}}$  = --0.43 at $V_{\mathrm{LC}}$ = --0.45 V. The spin-up current through the excited states is supposed to be small at this $V_{\mathrm{LC}}$, since only a small number of spin-up non-equilibrium electrons are transferred from EC$_{\mathrm{LL1}\uparrow}$ to EC$_{\mathrm{LL0}\uparrow}$. Therefore, the obtained negative value of $P_{\mathrm{S}}$ strongly indicates that spin-down electrons are preferentially injected into the QD. $P_{\mathrm{S}}$ is limited to be --0.43 supposedly because of the contribution of excited state conduction as is the case for the spin-up injection.

\section{Quantitative analysis of the contribution of excited states}

In Figs. 3(a) and (f) we observe the spin-dependent QD current vs. $V_{\mathrm{LC}}$ where the change in the QD current persists until the 2DEG beneath gate LC is completely depleted at $V_{\mathrm{LC}}$ = --0.59 V. On the other hand, for $B_{\perp}$ = 1.5 T, the depletion of spin-down electrons is expected to occur at $V_{\mathrm{LC}}$ = --0.55 V, because the difference between $V_{\alpha}$ and $V_{\beta}$ and also that between $V_{\delta}$ and $V_{\varepsilon}$ in Figs. 3(a) and (f) indicate that the spin splitting in a Landau level corresponds to the $V_{\mathrm{LC}}$ difference of 40 mV. Therefore, the QD current observed at the LL0 spin-down ridges for $V_{\mathrm{LC}}<-0.55$ V is supposedly the spin-up current through the excited states which we already described above. This contribution appears more clearly in the higher range of $B_{\perp}$. The QD current for the LL0 spin-up ridge of peak C, $I_{\uparrow}$, and that for the LL0 spin-down ridge of peak A, $I_{\downarrow}$, observed at $B_{\perp}$ = 1.75 T are shown by red and blue dashed curves, respectively, as a function of $V_{\mathrm{LC}}$ in Fig. 6. While $I_{\uparrow}$ shows distinct changes similar to that at $B_{\perp}$ = 1.5 T with $V_{\alpha}$ = --0.54 V, $V_{\beta}$= --0.49 V, and $V_{\delta}$ = --0.42 V, $I_{\downarrow}$ largely decreases with decreasing $V_{\mathrm{LC}}$ only at $V_{\mathrm{LC}}$ = --0.54 V. As well as the result for $B_{\perp}$ = 1.5 T shown in Fig. 3(a),  $V_{\alpha}$  and $V_{\delta}$  are assigned to the left edges of the $e^{2}/h$ and $3e^{2}/h$ plateaus at $B_{\perp}$ = 1.75 T. Consequently, the change in $I_{\downarrow}$ for $V_{\mathrm{LC}} < -0.54$ V also implies depopulation in the outermost spin-up EC and suggests that a spin-up current is the dominant component of $I_{\downarrow}$ in this $V_{\mathrm{LC}}$ range.

\begin{figure}
\includegraphics{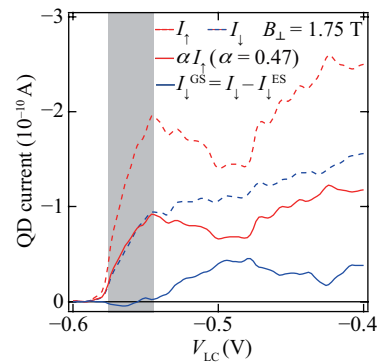}
\caption{(Color online)Pure spin-down QD current at the LL0 spin-down ridge ($I_{\downarrow}^{\mathrm{GS}}$), obtained by subtracting the spin-up contribution through the excited states ($\alpha I_{\uparrow}$) from the raw QD current at peak A ($I_{\downarrow}$), where $B_{\perp}$ = 1.75 T with $V_{\mathrm{S1}} = -60$ $\mu$V and  $V_{\mathrm{S2}} = 0$ $\mu$V for spin-up non-equilibrium ECs. $\alpha I_{\uparrow}$ is obtained by fitting the raw QD current at peak C ($I_{\uparrow}$) to $I_{\downarrow}$ in the $V_{\mathrm{LC}}$ range shown in gray. }%
\end{figure}

Based on this consideration, we evaluate the excess spin-up QD current through the excited states at the LL0 spin-down ridge. We simply assume an excess spin-up QD current $\alpha I_{\uparrow}$, in which $\alpha$ represents the efficiency of conduction through the excited states, to fit the $I_{\downarrow}$ data for --0.58 V $< V_{\mathrm{LC}} < V_{\alpha}$ (shown by the gray region in Fig. 6) with $\alpha$ as a fitting parameter. In this $V_{\mathrm{LC}}$ range, we assume that there is only spin-up excess current at the LL0 spin-down ridge, because $P_{\mathrm{S}}$ shows nearly its maximum value, suggesting the minimum spin-down QD current caused by complete depletion of spin-down electrons under gate LC. In the following, we neglect the conduction of spin-down current through the excited states to  $I_{\uparrow}$ at the LL0 spin-up ridge, because of the very weak coupling of the spin-down ECs to the QD. The obtained $\alpha I_{\uparrow}$ and the pure spin-down QD current through the ground state, $I_{\downarrow}^{\mathrm{GS}} = I_{\downarrow}-\alpha I_{\uparrow}$, are shown by the red and blue solid curves, respectively, in Fig. 6. $I_{\downarrow}^{\mathrm{GS}}$ shows a decrease at $V_{\mathrm{LC}} < -0.49$ V, indicating the depopulation of the outermost spin-down EC.

The efficiency $\alpha$ of conduction through the excited states influences the EC spin-filtering efficiency. As Fig. 4(b) shows, the EC spin-filtering efficiency decreases with increasing $B_{\perp}$ at high $B_{\perp}$. This is inconsistent with the expectation for the higher EC spin-filtering efficiency due to the larger spatial separation of spin-resolved ECs in higher $B_{\perp}$. This decrease of $P_{\mathrm{S}}$ was previously discussed in terms of an increase in a spin-up current through the excited states at the LL0 spin-down ridges.\cite{Hitachi} However, the origin has not yet been fully investigated. Figure 7(a) shows $\alpha$ obtained from the fittings as a funtion of $B_{\perp}$ for peaks A, B, C and D. We find $\alpha$ increases with $B_{\perp}$ for $B_{\perp}>$ 1.6 T. This result agrees with the above scenario which can explain the decrease in $P_{\mathrm{S}}$ with increasing $B_{\perp}$ in Fig. 4(b). The increase in $\alpha$ with decreasing $B_{\perp}$ for $B_{\perp}<$ 1.6 T might be due to the broadening of QD levels by large tunnel coupling in the presence of a weak magnetic confinement. This is also consistent with the larger $\alpha$ for peaks with a larger electron number at negatively small $V_{\mathrm{P}}$, which induces larger tunnel couplings.

\begin{figure}
\includegraphics{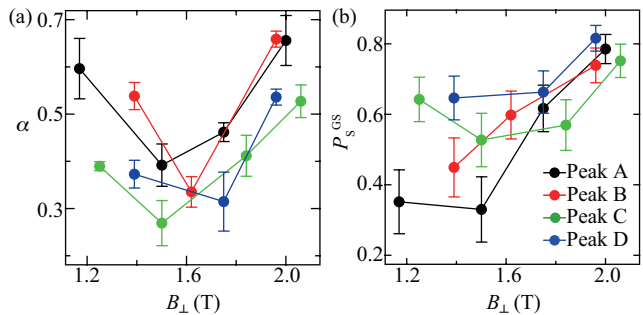}
\caption{(Color online)(a) The efficiency of transport through excited states $\alpha$ (a) and spin polarization $P_{\mathrm{S}}^{\mathrm{GS}}$ calculated using $I_{\uparrow}$ and $I_{\downarrow}^{\mathrm{GS}}$ (b) as a function of $B_{\perp}$ for peaks A, B, C, and D. }%
\end{figure}

We calculate the spin polarization $P_{\mathrm{S}}^{\mathrm{GS}}$ of the QD current through ground states alone, using $I_{\downarrow}^{\mathrm{GS}}$ instead of  $I_{\downarrow}$ in Eq.(1) at $V_{\mathrm{LC}} = -0.4$ V, where the outermost spin-up and spin-down ECs are in equilibrium. The contribution of the excess current through the excited states is approximately excluded in $P_{\mathrm{S}}^{\mathrm{GS}}$. This may reflect the EC spin-filtering efficiency more accurately than $P_{\mathrm{S}}$ obtained from the raw data of $I_{\uparrow}$ and $I_{\downarrow}$ in Fig. 4(b). Figure 7(b) shows the $P_{\mathrm{S}}^{\mathrm{GS}}$ as a function of $B_{\perp}$ for peaks A, B, C, and D. $P_{\mathrm{S}}^{\mathrm{GS}}$ is higher than the value evaluated from the raw experimental data shown in Fig. 4(b). Also, $P_{\mathrm{S}}^{\mathrm{GS}}$ monotonically increases with $B_{\perp}$, in contrast to $P_{\mathrm{S}}$ shown in Fig. 4(b). This trend agrees with the above expected $B_{\perp}$ dependence of EC spin-filtering efficiency, supporting the validity of our method to evaluate the contribution of the excited states discussed in Fig. 6. However, the assumption that there is only spin-up excess current at the LL0 spin-down ridge for --0.58 V $< V_{\mathrm{LC}} < V_{\alpha}$ means that $P_{\mathrm{S}} \approx 1$ for spin-up non-equilibrium ECs in this $V_{\mathrm{LC}}$ range. This is yet to be experimentally confirmed.

\section{Conclusion}
We have measured spin-dependent currents through a QD from spin-polarized non-equilibrium ECs generated by biasing a surface gate placed at a distance from the QD. The QD current shows spin-dependent changes with the gate voltage, indicating spin-resolved control of the local filling factor beneath the gate. We have electrically switched the spin polarization of the non-equilibrium ECs between spin-up and spin-down by changing the bias voltages applied to different source Ohmic contacts. As the result, the spin polarization of the QD current is switched as well, which is the demonstration of the selective injection of both spin-up and spin-down single electrons into a QD. Also, the QD current shows the spin-independent component attributed to the electron conduction through the QD excited states. Through the numerical analysis, we have rectified the underestimation of the EC spin-filtering efficiency by eliminating the contribution of the excited states. The obtained EC spin-filtering efficiency is higher than that evaluated from the raw experimental data, and increases with a magnetic field as expected with the increase in the spatial separation between ECs.

The selective spin injection into QDs from ECs may be important not only in the development of spin qubits in QDs for quantum computing but also for connections between local qubits in QDs and flying qubits in ECs.\cite{Stace} With the help of spin manipulations of electrons traveling in ECs that have been recently reported,\cite{Karmakar,Nakajima} it may enable one to inject arbitrary electron spin states into QDs, which may be utilized for quantum memory or quantum state tomography.

\section{Acknowledgements}
This work was supported by Grants-in-Aid for Scientific Research A (No. 25246005) and S (No. 26220710), Innovative Areas "Quantum Cybernetics" (No. 21102003) and “Nano Spin Conversion Science” (No. 26103004), Funding Program for World-Leading Innovative R$\&$D on Science and Technology (FIRST), Intelligence Advanced Research Projects Activity project "Multi-Qubit Coherent Operations" through Copenhagen University, MEXT Project for Developing Innovation Systems, QPEC, The University of Tokyo, The Strategic Information and Communications R$\&$D Promotion Programme (SCOPE) of the Ministry of Internal Affairs and Communications Government of Japan (MIC), and ImPACT Program of Council for Science, Technology and Innovation (Cabinet Office, Government of Japan).



%

%


\bibliography{NoneqSpinInjectionFinal2.bbl}

\end{document}